\documentstyle[aps,preprint,epsfig]{revtex}

\begin{document}

\draft


\title{A simple model for the metal-insulator transition in a two-dimensional
electron gas}

\author{J.\ C.\ Flores$^{\dag}$, V.\ Bellani and
F.\ Dom\'{\i}nguez-Adame$^{\ddag}$}

\address{INFM-Dipartimento di Fisica ``A.\ Volta", Universit\'{a}
di Pavia,  I-27100 Pavia, Italy}

\date{\today}

\maketitle

\begin{abstract}

We introduce an elementary model for the electrostatic self-consistent
potential in a two-dimensional electron gas. By considering the perpendicular
degree of freedom arising from the electron tunneling out of the system plane,
we predict a threshold carrier density above which this effect is relevant. The
predicted value agrees remarkably well with the onset for the insulator to
quasi-metallic transition recently observed in several experiments in
SiO$_2$--Si and AlGaAs--GaAs heterojunctions. 

\end{abstract}


Anderson transition~\cite{Anderson58} in disordered solids still raises great
interest among researchers. While metal-insulator transition (MIT) is well
established in three-dimensional (3D) disordered systems, the situation is not
completely understood in one- (1D) and two-dimensions (2D). Several decades
ago, a number of papers yields the general belief that in those systems all
eigenstates are exponentially localized and a MIT no longer
exists\cite{Mott61,Abrahams79,Lee85}. Although this seems to be correct in most
low-dimensional systems, there exist known exceptions to this rule. Thus, it
has been found that extended states may appear in 1D random systems upon
introducing either 
short-range~\cite{Flores1,Dunlap,Wu,Phillips91,Flores2,PRBKP} or 
long-range~\cite{Moura98,Izrailev99} correlations in the disorder. These purely
theoretical considerations were put forward for the explanation  of high
conductivity of doped polyaniline~\cite{Phillips91} as well as transport
properties of random semiconductor superlattices~\cite{Bellani99}.

The Anderson transition induced by diagonal disorder at the band center in
finite 2D systems was already studied by Yoshino and Okazaki~\cite{Yoshino77}.
The relevance of these midgap states for the metallic conductance of 2D systems
has been discussed in detail by Licciardello and
Thouless~\cite{Licciardello78}. These authors suggested that between the
mobility edges there is a tendency for the conductance to decrease slowly as
the sample size is increased and that they may be no absolute minimum metallic
conductance. However, recent experiments have provided clear evidences of the
MIT-like in high-quality 2D electron and hole systems. After the pioneering
work by Kravchenko~{\em et al.\/} on electrons in
Si~\cite{Kravchenko94,Kravchenko95,Kravchenko96} and more recently by
Hanein~{\em et al.\/} on holes in GaAs~\cite{Hanein98}, it has be become clear
that 2D gases undergo a crossover from an insulating regime at low density to a
metalliclike behavior at high density, where the quasi-metallic phase is
characterized by a strong decrease of the resistivity as the temperature
decreases. Discrepancies between the standard one-parameter scaling
theory~\cite{Abrahams79}, establishing that the 2D gas should be insulating, 
and the above mentioned experiments are usually attributed to a strong
electron-electron interaction. A number of theoretical models have then been
proposed to explain the observed transition in 2D systems, ranging from new
liquid phases~\cite{He98}, Wigner glass~\cite{Chakravarty98}, spin-orbit
induced transition~\cite{Pudalov97}, decoherence due to quantum
fluctuations~\cite{Mohanty99}, superconducting phase~\cite{Phillips97} and
anyon superconducting model~\cite{Zhang97}. The basic ingredient of these
models is the assumption that the system is purely 2D and, consequently, new
phenomena are to be considered to explain the observed transition. In this work
we undertake a different way by considering the electronic motion in the
perpendicular direction. In so doing, we find the conditions under which this
degree of freedom could be relevant. Surprisingly, the critical carrier density
leading to perpendicular motion is close to that determined in experiments to
observe the transition. Our main aim is then to point out the importance of
this perpendicular degree of freedom, which should be taken into account in
more elaborated models.

Our approach is based on the competition between the confining potential,
appearing at the heterojunction even at zero gate voltage, and the repulsive
potential arising from the excess carrier at nonzero gate voltage. To proceed,
let us start by considering the heterojunction at zero gate voltage. Since we
are only interested in the basic phenomena without entering in many details, we
make use of a simple variational Hartree calculation presented in standard
textbooks~\cite{Davies98}. This will make our reasonings clearer while keeping
a good qualitative description of the involved physics. The envelope function 
of the lowest subband of a 2D electron gas is reasonably well accounted for by 
the Fang-Howard trial function:
\begin{equation}
\chi(z)=\sqrt{\frac{b^3}{2}}\,z\,\exp\left(-{1\over 2}\,bz\right),
\label{FH}
\end{equation}
where $z$ is the coordinate along the growth direction (perpendicular to
the heterojunction) and $b$ is determined by minimizing the kinetic energy 
plus the Hartree energy per electron (see Ref.~\cite{Davies98} for details).
In terms of the effective Bohr radius, $a^{*}$, and the 2D electron density,
$n^{\mathrm o}_s$, the variational parameter is roughly given by
\begin{equation}
b=\left( {16\pi n^{\mathrm o}_s\over a^{*}}\right)^{1/3}, 
\label{parameter}
\end{equation}
while the maximum value of the Hartree potential is~\cite{Davies98} 
\begin{equation}
V_0=3\,\frac{e^2n^{\mathrm o}_s}{\epsilon\, b},
\label{vmax}
\end{equation}
$\epsilon$ being the dielectric constant of the medium.
The Hartree potential increases smoothly from zero up to $V_0$ on increasing
the distance $z>0$ from the heterojunction. Thus, at zero gate voltage the
electrons lie on the lower subband and confined, thus forming a 2D gas. The
lower subband energy is expressed as:
\begin{equation}
\varepsilon_1=2.5\,\frac{e^2n^{\mathrm o}_s}{\epsilon\, b}.
\label{epsilon}
\end{equation}

\begin{figure}
\centerline{\epsfig{file=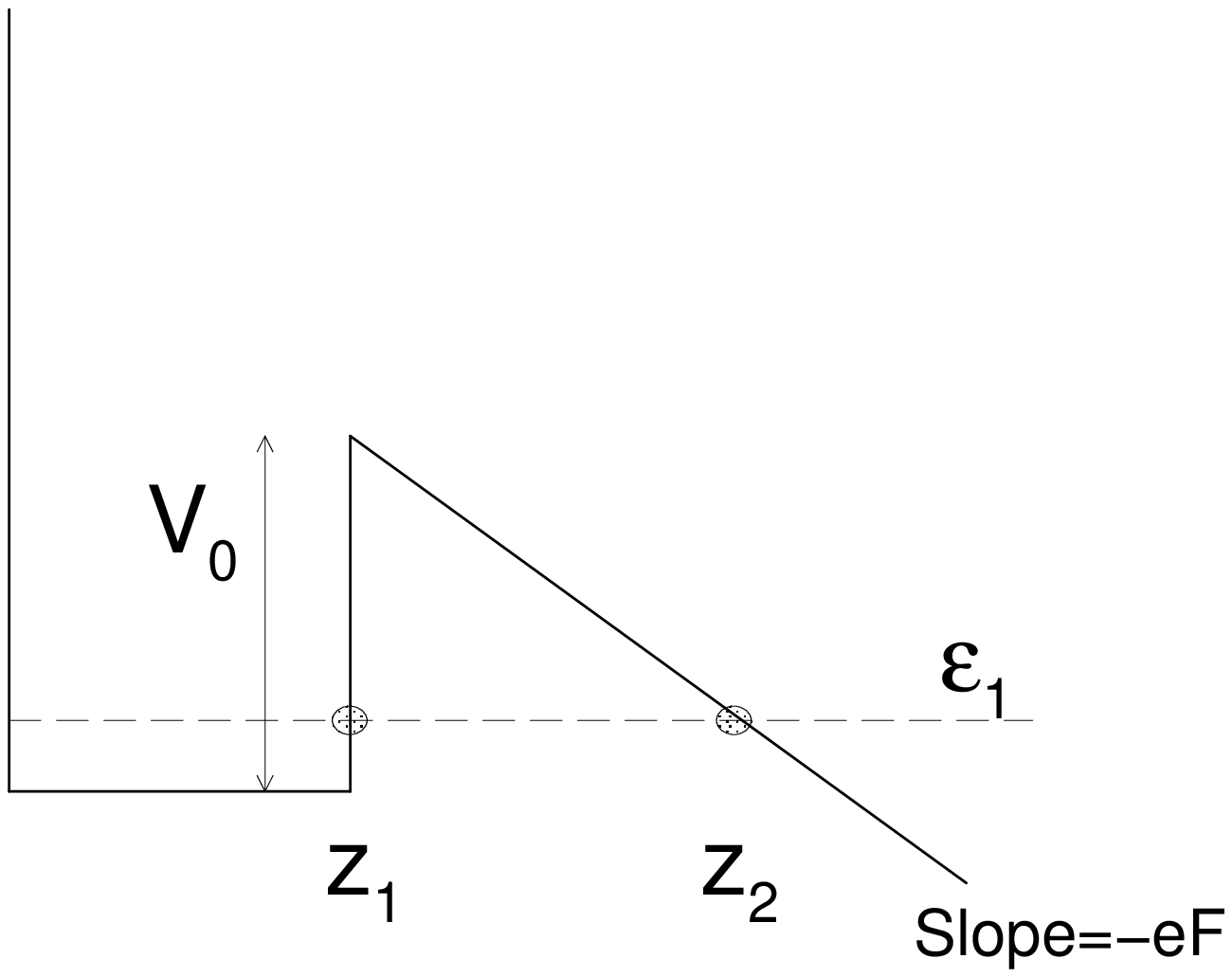,width=7.0cm}}\vspace{5mm}
\caption{Schematic diagram of the band-edge profile of the heterojunction
at high carrier density.}
\label{fig1}
\end{figure}

The situation may be different when the gate voltage induces an excess carrier
density $\Delta n_s$, as we will show below. This excess carrier is not
compensated by any other charge close to the heterojunction, thus leading to a
local negative charge density. As a crude and first approximation, we assume
that this excess charge density is confined to a plane close to the
heterojunction. By solving the Poisson equation, one can obtain the potential
energy due to this charged plane, namely $-eFz$ for $z>0$, where
\begin{equation}
F={e\Delta n_s\over 2\epsilon}.
\label{F}
\end{equation}
Due to this repulsive potential, electrons may tunnel through the barrier
formed by the Hartree potential plus $-eFz$ and escape. To account for the
tunneling process, we replace the actual potential by that depicted in
Fig.~\ref{fig1}. This replacement is not essential in the calculations since
both the confining  Hartree potential and the repulsive potential due to the
charged plane are known, but it allows us to obtain closed analytical
expressions for the transmission coefficient. Since the Fermi energy is usually
very small in this system, we can also safely neglect it when calculating the
transmission coefficient and take $E=\varepsilon_1$, as indicated in
Fig.~\ref{fig1}. Thus, the classical turning points are $z_1$ and
$z_2=z_1+(V_0-\varepsilon_1)/eF$, shown in the figure. From the WKB
semi-classical approximation, the transmission coefficient is given by 
\begin{eqnarray}
T&=&\exp\Bigg[-2\sqrt{2m^*\over \hbar^2} \int_{z_1}^{z_2}\>dz\,
\sqrt{V_0-\varepsilon_1-eF(z-z_1)} \Bigg] \nonumber \\
&=&\exp\Bigg[-{4\over 3}\sqrt{2m^*\over \hbar^2}\,
{(V_0-\varepsilon_1)^{3/2}\over eF}\Bigg].
\label{transmission}
\end{eqnarray}
The onset for the crossover from a 2D to 3D behavior can be determined from the
condition that the transmission probability is large. We then assume that this
crossover appears when the exponent in~(\ref{transmission}) is of the  order of
unity in absolute value. This occurs for a critical density $n_s^c=\Delta
n_s+n^{\mathrm o}_s$ such that
\begin{equation}
{4\over 3}\sqrt{2m^*\over \hbar^2}\,{(V_0-\varepsilon_1)^{3/2}\over eF} \sim 1.
\label{condition}
\end{equation}
Since $F$ is a function of $\Delta n_s$, we can readily determine the value of
the critical density from the above condition for which the probability for
tunneling out of the well is large. Using~(\ref{vmax}), (\ref{epsilon})
and~(\ref{F}) we finally obtain $n_s^c \sim (5/3)n^{\mathrm o}_s$. Since the
carrier density at zero gate voltage is related to the Fermi energy
$E_F^{\mathrm o}$ by the expression $n^{\mathrm o}_s=2m^*E_F^{\mathrm o}/h$, 
we finally arrive at the condition
\begin{equation}
n_s^c \sim 7\,E_F^{\mathrm o}\,m^* \times 10^{11}\,{\mathrm cm}^{-2},
\label{final}
\end{equation}
where the Fermi energy is measured in meV and the effective mass in units of
the free electron mass.


Now let us compare our qualitative prediction~(\ref{final}) with the
experimental values. A 2D electron gas in Si has been studied by Kravchenko
{\em et al\/}.~\cite{Kravchenko96}, who observed the transition for an electron
density $0.85\times 10^{11}\,$cm$^{-2}$. The Fermi energy was $E_F=0.6\,$meV
and the effective mass $m^*=0.2m$. Inserting both values in~(\ref{final}) we
obtain a critical density $n_s^c \sim 0.84 \times 10^{11}\,$cm$^{-2}$. On the
other side, 2D hole gas in GaAs was demonstrated to undergo a MIT at a hole
density $0.15\times 10^{11}\,$cm$^{-2}$ by Hanein~{\em et
al.\/}~\cite{Hanein98}. Taking the values $E_F=0.04\,$meV and $m^*=0.4m$ we get
from~(\ref{final}) that $n_s^c \sim 0.11 \times 10^{11}\,$cm$^{-2}$. The
agreement should be regarded as surprisingly good, in view of the crude
approximation we made to obtain it. As a conclusion, our model points out the
relevance, under some circumstances, of the perpendicular degrees of freedom in
the so-called 2D electron gases. As soon as the electron gas becomes a 
non-perfect 2D system, the scaling theories predicts the occurrence of a MIT
transition like that recently observed. 

The authors warmly thank I.\ G\'{o}mez for his useful comments and criticisms,
and E.\ Diez, C.\ Kanyinda-Malu and M.\ Hilke for helpful conversations. FDA
was supported by CAM under Project~07N/0034/98. JCF was supported by CICOPS
fellowship of University of Pavia.

\end{document}